\documentclass[aps,showpacs,twocolumn, tightenlines]{revtex4}
\usepackage{epsfig,amssymb,bm,graphics}
  \usepackage{amsbsy}

\begin{document}


\title{
Selected vector-meson decay-distributions in reactions of
polarized photons
with protons\\
}

  \author{A.I.~Titov$^{a,b}$ and B.~K\"ampfer$^{a,c}$}
  \affiliation{
  $^a$Forschungzentrum Dresden-Rossendorf, 01314 Dresden, Germany\\
  $^b$Bogoliubov Laboratory of Theoretical Physics, JINR,
   Dubna 141980, Russia\\
  $^c$ Institut f\"ur Theoretische Physik, TU~Dresden, 01062 Dresden,
  Germany
  }


\begin{abstract}
  We develop a formalism for studying vector
  meson ($V$) photo-production at the proton ($p$) with
  polarized photons,
  $\vec\gamma p \to V p $, 
  through an analysis of the
  decay distribution in the channel $V\to\pi^0\gamma$.
  We show that this decay distribution
  differs noticeably from the distributions of purely
  hadronic decays, like $\phi\to K^+K^-$, $\omega\to \pi^0\pi^+\pi^-$.
  Formulas for the decay distributions are presented which are suitable
  for data analysis and interpretation.
\end{abstract}

  \pacs{13.60.Le, 13.20.Jf, 13.25.Jx}

  \maketitle

   Photo-production of iso-scalar vector mesons ($V$) at the proton ($p$),
   $\gamma p \to V p $,
   plays an important role
   in understanding phenomena of hadronic physics
   which are far beyond the scope of perturbative QCD.
   Thereby, reactions with polarized photons ($\vec \gamma$)
   at relatively low energies are particularly interesting.
   For example, photo-production of $\phi$ mesons
   at forward angles can provide some information on the Pomeron
   exchange channel.
   A recent analysis of the LEPS collaboration
   shows a sizeable deviation
   from predictions based on conventional approaches~\cite{TL03},
   in particular, as the data show a bump structure
   at photon energies $E_\gamma\sim 2$~GeV~\cite{Mibe05}.
   Another peculiarity of the LEPS data is a strong deviation
   of the spin-density matrix element $\rho^1_{1-1}$ from 0.5,
   which points to a sizable contribution of un-natural parity
   exchange processes, contrary to (conventional) expectations based on the
   dominance of the Pomeron exchange channel.

   $\omega$ photo-production near the threshold provides further
   unique information on the mechanism of nucleon
   resonance ($N^*$) excitation~\cite{Ajaka06,TL02,Zhao01,Shklyar04}
   and the strength of the $\omega NN^*$ coupling which are
   crucial for understanding the structure of baryon
   resonances~\cite{Capstick00}.

   Most experiments of $\phi$ and $\omega$ photo-production
   are aimed at analyzing the angular distribution of the exclusively outgoing
   hadrons in reactions with polarized photons.
   The corresponding formalism for such analysis,
   expressing the angular distributions of the outgoing hadrons
   in terms of the spin-density matrix elements, is developed in
   Ref.~\cite{Schilling70}. However, there is also an opportunity
   to extend the study of vector meson photo-production
   by measuring the $V \to \pi^0\gamma$ decay channel.
   The Crystal Barrel detector~\cite{Aker}, which is now operated
   at the electron accelerator ELSA at Bonn and
   delivers a new data complementary to the previous ones,
   providing independent information about
   various aspects of $\phi$ and $\omega$ photo-production~\cite{Schmiden}.

   Given this motivation,
   the aim of our Note is to extend the formalism of
   Ref.~\cite{Schilling70} to the case of the $V\to\pi^0\gamma$
   decay distribution.
   We are going to present useful formulas for corresponding distributions
   and asymmetries. Our focus is on iso-scalar vector meson photo-production
   off the proton.

Taking into account the small total decay width of $\omega$ and
$\phi$ mesons, one can express the cross section of the reaction
$\vec \gamma p\to Vp\to \pi^0\gamma p$ with linearly polarized
photons as
  \begin{eqnarray}
\frac{d\sigma^{\pi^0\gamma}(\Omega_\pi,\Psi)}{dt d\Omega_\pi} =
B_{\pi^0\gamma}\frac{d\sigma_0}{dt}\,
W^{\pi^0\gamma}(\Omega_\pi,\Psi)~, \label{E1}
\end{eqnarray}
where $\Psi$ is the angle between the polarization of the incoming
photon, ${\bm \varepsilon}=(\cos\Psi,\sin\Psi, 0)$, and the
production plane; $\Omega_\pi$ denotes the solid angle of the
momentum of the outgoing pion in the vector meson rest frame,
$W^{\pi^0\gamma}(\Omega_\pi,\Psi)$ is the decay distribution; $
B_{\pi^0\gamma}={\Gamma_{\pi^0\gamma}}/{\Gamma_{\rm tot}}$ stands
for the branching ratio of the $V\to\pi^0\gamma$ decay, and
${d\sigma_0}/{dt}$ is the differential cross section of the vector
meson production for an unpolarized photon beam. Note, that our
consideration is valid for $V\to\eta\gamma$ decay as well, and in
particularly for $\phi\to\eta\gamma$ decay mode, which branching
ratio exceeds the branching ratio of $\phi\to\pi^0\gamma$ decay by
order of magnitude. But for simplicity, we will not distinguish
between these two modes, assuming that the result for
$V\to\eta\gamma$ decay is obtained by the evident substitution
$\pi^0\to\eta$.

For calculating the decay distribution we cast the decay amplitude
in the form
\begin{eqnarray}
T^{V\to\pi^0\gamma} =\frac{g_{V\pi^0\gamma}}{M_{V}}
\epsilon^{\mu\nu\alpha\beta} V_\mu \,
\varepsilon^{*\,(\gamma)}_\alpha(\lambda_{(\gamma)}) \, q_\nu \,
k_{\beta}~, \label{E2}
\end{eqnarray}
where $V_\mu$ and $\varepsilon^{*\,(\gamma)}_\alpha$ are the
polarization four-vectors of the vector meson $V$ with momentum
$q$ and mass $M_V$ and the outgoing photon $\gamma$ with momentum
$k$, respectively; $g_{V\pi^0\gamma}$ is the coupling strength of
the $V\pi^0\gamma$ interaction. By summing up the polarization
states $\lambda_{(\gamma)}$ of the outgoing photon in the total
probability for the reaction $\vec \gamma p\to \pi^0\gamma p$, one
gets the following expression for the decay distribution in the
vector meson's rest frame
\begin{eqnarray}
W^{\pi^0\gamma}(\Omega_\pi,\Psi) =\frac{3}{8\pi}\frac{1}{N} \left(
{\vec {\cal M}}{\vec {\cal M}}^* -({\vec {\cal M}}\cdot{\vec
n}_\pi) ({\vec {\cal M}}\cdot{\vec n}_\pi)^* \right), \label{E3}
\end{eqnarray}
where ${\vec n}_\pi$ is the unit three-vector along direction of
flight of the outgoing $\pi$ meson,  $\vec {\cal M}$ is the
amplitude for the vector meson photo-production with components
${\cal M}_\lambda$ referring to the three polarization states
$\lambda=\pm1,0$. Thus,
\begin{eqnarray}
{\vec {\cal M}}\cdot{\vec n}_\pi=\sqrt{\frac{4\pi}{3}}
\sum\limits_{\lambda} {\cal M}_{\lambda}
Y_{1\lambda}(\Omega_\pi)~. \label{E4}
\end{eqnarray}
The normalization factor $N$ in Eq.~(\ref{E3}) is related to the
differential cross section of the vector meson photo-production by
an unpolarized photon beam
\begin{eqnarray}
\frac{d\sigma_{0}}{dt} =\frac{1}{16\pi(s-M_p^2)^2}\,|N|^2~.
\label{E5}
\end{eqnarray}
Here, $s$ is the Mandelstam variable for the entrance channel
$\gamma p$, and $M_p$ denotes the proton mass. It is worth noting
that the distribution for purely hadronic decays, say $\phi\to
K^+K^-$ or $\omega\to\pi^+\pi^-\pi^0$, differs noticeably from
Eq.~(\ref{E3}):
\begin{eqnarray}
W^{h}(\Omega_\pi,\Psi) =\frac{3}{4\pi}\frac{1}{N} ({\vec {\cal
M}}\cdot{\vec n}_h) ({\vec {\cal M}}\cdot{\vec n}_h)^*~,
\label{E6}
\end{eqnarray}
where ${\vec n}_h$ is the unit three-vector along direction of
flight of a kaon in case of the $\phi$ meson decay; for the
$\omega$ meson decay, ${\vec n}_h$ defines the direction of the
vector ${\bf p}_{\pi^0}\times({\bf p}_{\pi^+}-{\bf p}_{\pi^-})$ .

The evaluation of Eq.~(\ref{E3}) for polarized photons is
performed with the photon density matrix~\cite{Schilling70}
\begin{eqnarray}
\rho(\gamma)=\frac12 I +\frac12{\bf P}_\gamma\cdot{\bm \sigma}~,
\label{E7}
\end{eqnarray}
where $I$ is the $2\times2$ unity matrix, and $\sigma_i$ stand for
the Pauli matrices. The three components of ${\bf P}_\gamma$ read
for linear and circular polarizations
\begin{equation}
{\bf P}_\gamma = \left\{
\begin{array}{ll}
P_\gamma\left(-\cos2\Psi,-\sin2\Psi,0\right)& ({\rm lin. \, pol.}),\\[1mm]
P_\gamma\left(0,0,\pm 1\right) & ({\rm circ. \,pol.}),
\end{array}
\right.  \label{E8}
\end{equation}
where $P_\gamma$ is the degree of the polarization. The final
result for linearly polarized photons, expressed in terms of the
spin-density matrix elements, reads
\begin{eqnarray}
  W^{\pi^0\gamma}(\Omega_\pi,\Psi) &=&
  W^{\pi\gamma}_0(\Omega_\pi)
  -P_\gamma\,W^{\pi^0\gamma}_1(\Omega_\pi)\cos2\Psi \nonumber\\
  &-&P_\gamma\,W^{\pi^0\gamma}_2(\Omega_\pi)\sin2\Psi~, \label{E9}
\end{eqnarray}
with
\begin{eqnarray}
&&W^{\pi^0\gamma}_0(\Omega_\pi) = \frac{3}{8\pi} \left(
1-\rho^0_{11}\sin^2\Theta-\rho^0_{00}\cos^2\Theta\right. \label{E101}\\
&&\quad\quad + \left.\rho^0_{1-1}\sin^2\Theta\cos2\Phi
+\sqrt{2}{\rm Re}\rho^0_{10}\sin2\Theta\cos\Phi
\right), \nonumber\\
&&W^{\pi^0\gamma}_1(\Omega_\pi) = \frac{3}{8\pi} \left(
2\rho^1_{11}+(\rho^1_{00}-\rho^1_{11})\sin^2\Theta\right. \label{E102}\\
&&\quad\quad + \left. \rho^1_{1-1}\sin^2\Theta\cos2\Phi
+\sqrt{2}{\rm Re}\rho^1_{10}\sin2\Theta\cos\Phi
\right), \nonumber\\
&&W^{\pi^0\gamma}_2(\Omega_\pi) = -\frac{3}{8\pi} \left( {\rm
Im}\rho^2_{1-1}\sin^2\Theta\cos2\Phi\right.  \label{E103}\\
&&\quad\quad + \left. \sqrt{2}{\rm {\rm
Im}}\rho^2_{10}\sin2\Theta\cos\Phi \right)~, \nonumber
\end{eqnarray}
while the decay distribution for circular polarized photons with
polarization $\lambda_\gamma=\pm1$ is
\begin{eqnarray}
W^{\pi^0\gamma}_{\lambda_\gamma}(\Omega_\pi) &=& W^{\pi^0\gamma}_0
- P_\gamma \frac{3\lambda_\gamma}{8\pi} \left(
  {\rm Im}\rho^3_{1-1}\sin^2\Theta\cos2\Phi\right.
  \nonumber\\
  &+&\,\left.
\sqrt{2}{\rm Im}\rho^3_{10}\sin2\Theta\cos\Phi \right)~.
\label{E11}
\end{eqnarray}
The spin-density matrix elements are defined by the components of
$\vec {\cal M}$; their explicit form may be found in
Ref.~\cite{Schilling70}.

\begin{figure}[hb!]
    \includegraphics[width=0.45\columnwidth]{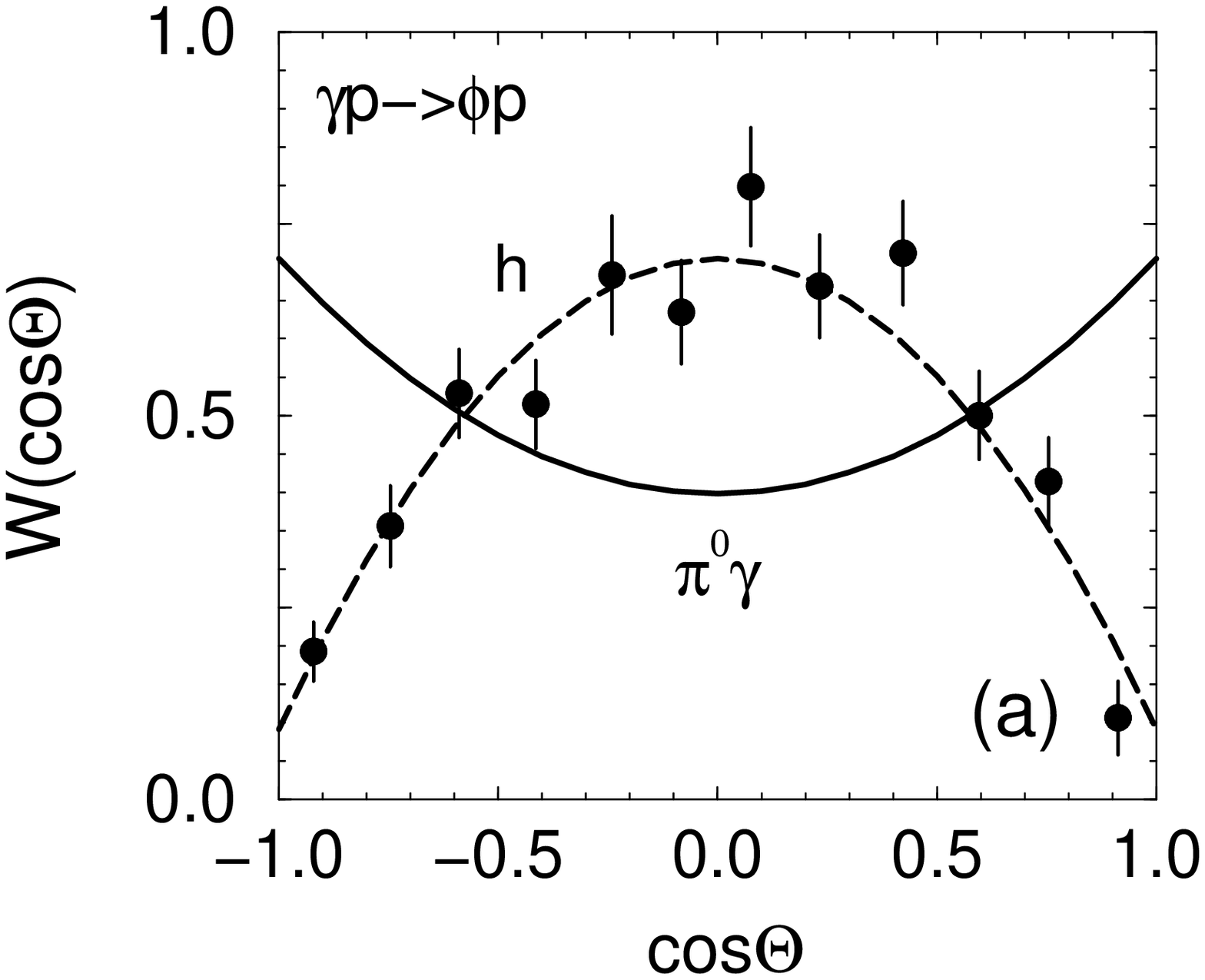}\qquad
    \includegraphics[width=0.45\columnwidth]{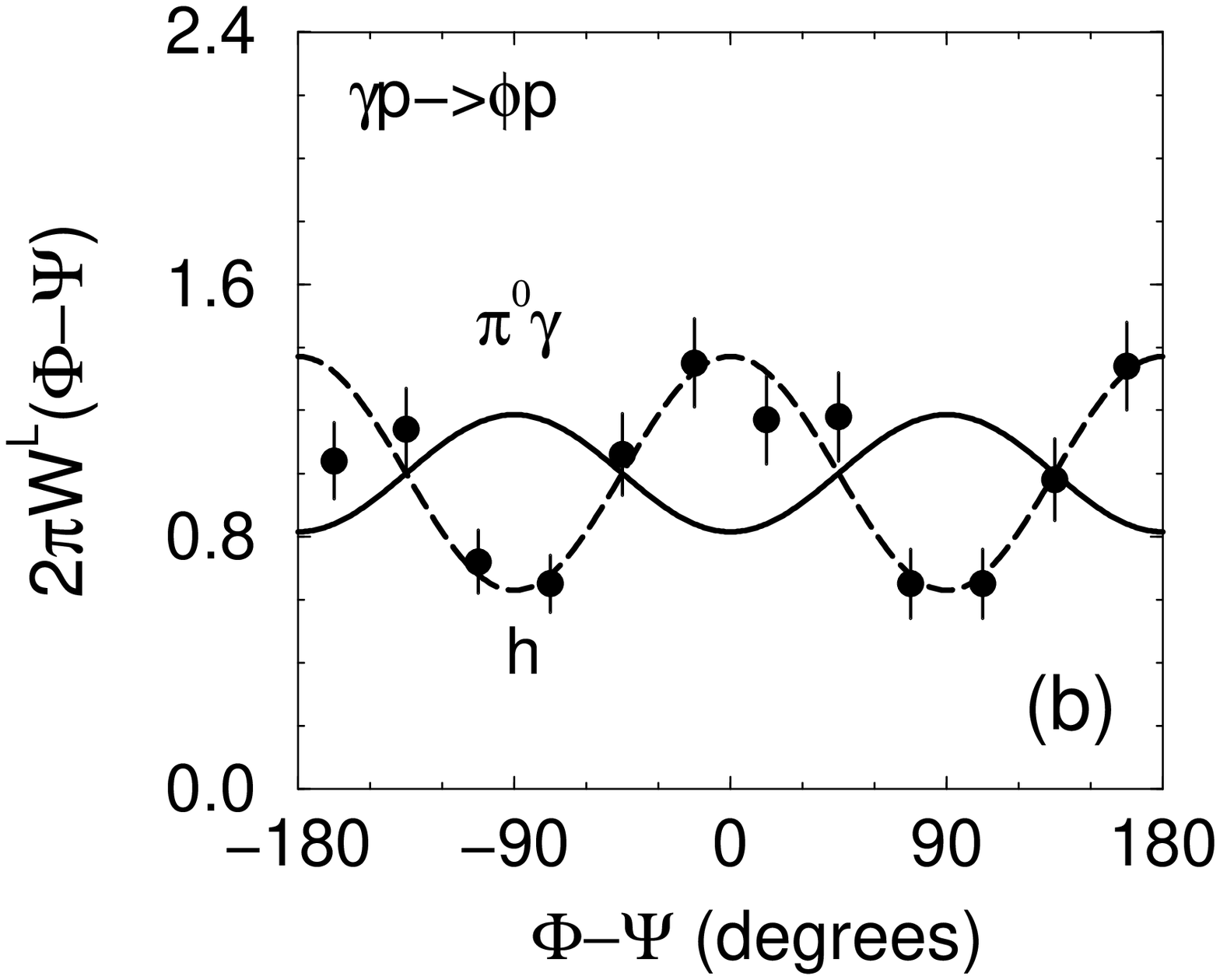}
    \caption{\small{
    The polar angular distribution (left panel) and
    the azimuthal angle decay distribution (right panel)
    for the $\vec \gamma p\to \phi p\to \pi^0\gamma p$ reaction
    (solid curve) and for the $\vec \gamma p\to \phi p\to K^+K^- p$ reaction
   (dashed curve).
    in the Gottfried-Jackson frame for $E_\gamma=1.97-2.17$~GeV and
    $t=t_{\rm max}-0.2$~GeV$^2$.
    The experimental data are taken from~\protect\cite{Mibe05}.
     \label{Fig:1}}}
    \end{figure}

Eqs.~(\ref{E101} - \ref{E11}) show that the angular distributions
for the $V\to\pi^0\gamma$ decay differ from the distributions of a
purely hadronic decay emerging from (\ref{E6})
\cite{TK07,TL03,Schilling70}. Thus, the polar angular
distributions in reaction with unpolarized photons, integrated
over the azimuthal angle $\Phi$, read for the $\pi^0 \gamma$ and
exclusively hadronic decays
\begin{eqnarray}
W^{\pi^0\gamma}(\Theta) &=& \frac{3}{8} \left(1+\cos^2\Theta +
\rho^0_{00}(1-3\cos^2\Theta) \right), \label{E12}\\
W^{h}(\Theta) &=& \frac{3}{4} \left(\sin^2\Theta -
\rho^0_{00}(1-3\cos^2\Theta) \right)~. \label{E13}
\end{eqnarray}
We emphasize the difference between the two distributions.

The azimuthal angle distribution, integrated over the polar angle
$\Theta$, for the two considered cases may be written in the
universal form
\begin{eqnarray}
2\pi W^{f}(\Phi,\Psi)&=& 1 -\Sigma^f_{\Phi}\cos2\Phi
-P_\gamma\Sigma^f_b\cos2\Psi\nonumber\\
  &+&P_\gamma\Sigma^f_d\cos2(\Phi-\Psi)~,
  \label{E14}
\end{eqnarray}
where the symbol $f$ labels either the $\pi^0\gamma$ or the pure
hadronic ($h$) decay; $\Sigma^f_{\Phi}$ denotes the azimuthal
decay asymmetry for unpolarized photon beam, $\Sigma^f_{b}$
denotes the beam asymmetry and $\Sigma^f_d$ is the azimuthal decay
asymmetry relative to the photon polarization plane. For
circularly polarized photons, the azimuthal angle distribution
reads
\begin{eqnarray}
2\pi W^{f}_{\lambda_\gamma}(\Phi,\Psi)= 1 + \lambda_\gamma
P_\gamma\Sigma^f_C\cos2\Phi~,
  \label{E15}~
\end{eqnarray}
where $\Sigma^f_C$ is the azimuth angle asymmetry for circularly
polarized photons with polarization $\lambda_\gamma=\pm1$. The
asymmetries are expressed through the spin-density matrices
  \begin{eqnarray}
\Sigma^{\pi^0\gamma }_{\Phi}&=&
-\frac12\Sigma^{h}_{\Phi}=-\rho^0_{1-1}~,\label{16}\\
\Sigma^{\pi^0\gamma }_{b}&=&
\Sigma^{h}_{b}=2\rho^1_{11}+\rho^1_{00}~,\label{17}\\
\Sigma^{\pi^0\gamma }_{d}&=&
-\frac12\Sigma^{h}_{d}=-\rho^1_{1-1}~.\label{18}\\
\Sigma^{\pi^0\gamma }_{C}&=& -\frac12\Sigma^{h}_{c}=-{\rm Im
}\rho^3_{1-1}~,\label{19}
\end{eqnarray}
where we assume $\rho^1_{1-1} \approx -{\rm Im}\rho^2_{1-1}$
according to \cite{TL03}. It is clear that the beam asymmetry does
not depend on the decay mode. However, the other asymmetries
depend on the decay mode: (i) they have opposite signs and (ii)
the absolute values for the purely hadronic decay is two times
larger.
\begin{figure}[ht!]
    \includegraphics[width=0.45\columnwidth]{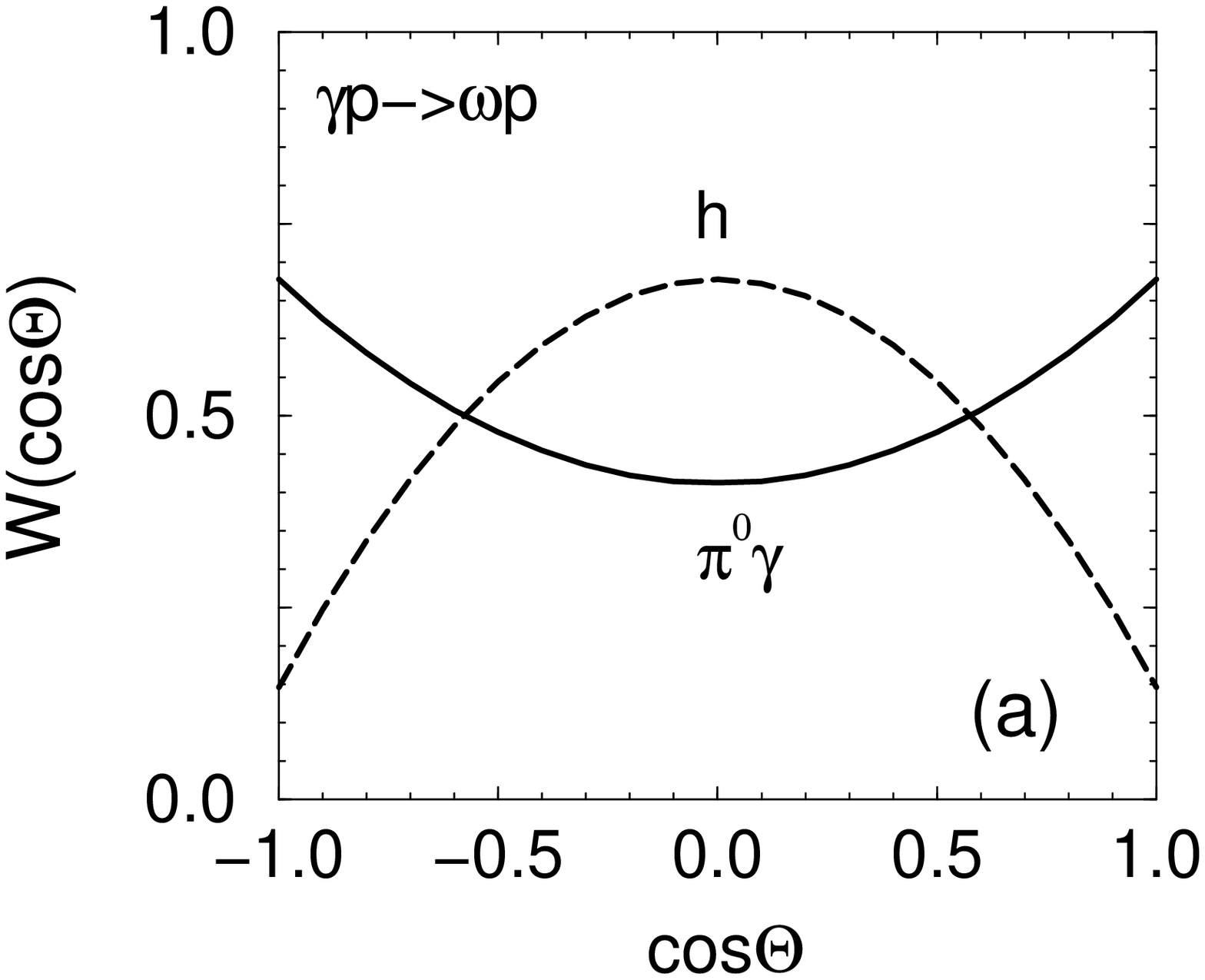}\qquad
    \includegraphics[width=0.45\columnwidth]{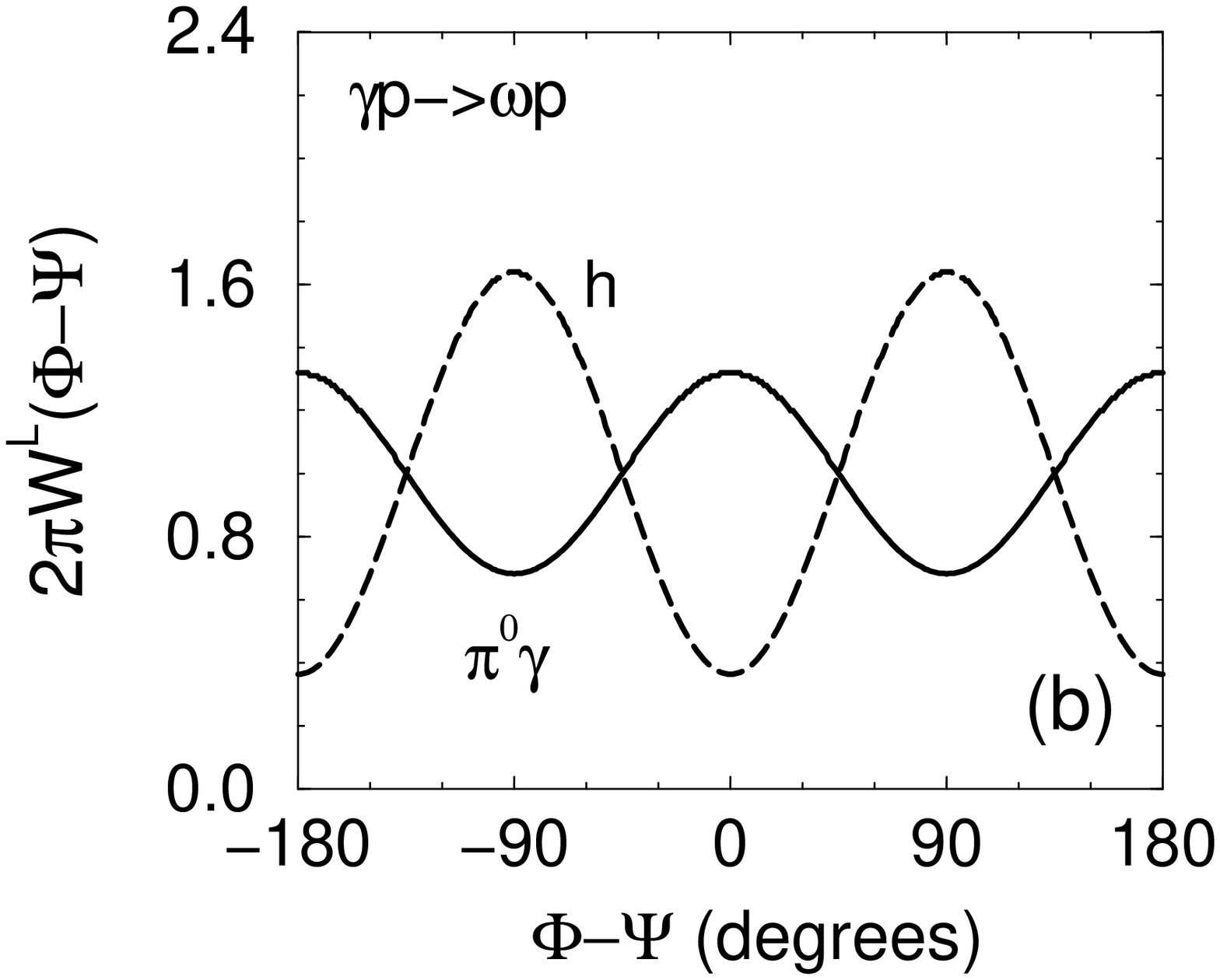}
    \caption{\small{The same as in Fig.~\protect\ref{Fig:1}
    but for $\omega$ meson photo-production.
    Our prediction is for $E_\gamma=1.6$~GeV and
    $t=t_{\rm max}-0.2$~GeV$^2$.
     \label{Fig:2}}}
    \end{figure}

The different distributions, for the case of $\phi$ meson
photo-production, are exhibited in Fig.~\ref{Fig:1}.  The left
panel shows the polar angular distribution according to
Eq.~(\ref{E13}). The spin-density matrix elements are calculated
in the Gottfried-Jackson system using the model of
Ref.~\cite{TK07}. The photon energy is integrated over the
interval $E_\gamma=1.97-2.17$~GeV, and the momentum transfer is
$t=t_{\rm max}-0.2$~GeV$^2$, where $t_{\max}$ corresponds to the
$\phi$ meson production at forward angle, i.e., $\theta=0$. The
experimental data are from~Ref.~\cite{Mibe05}. The azimuthal angle
distribution relative to the photon polarization plane is
exhibited in the right panel. One can see the striking difference
of the angular distributions for $K^+K^-$ and $\pi^0\gamma$
decays.

A similar comparison for $\omega$ meson photo-production is
presented in Fig.~\ref{Fig:2}. Here, the spin-density matrix
elements are calculated using the model of Ref.~\cite{TL02}. The
calculation is for $E_\gamma=1.6$~GeV and $t=t_{\rm
max}-0.2$~GeV$^2$. Again, there is a striking difference in the
angular distributions for the two considered decay modes.

In summary, we derived suitable formulas of the angular decay
distributions for $\phi$ and $\omega$ meson photo-production and
subsequent decay to $\pi^0\gamma$ ($\eta\gamma$). We found that
the distributions differ from corresponding distributions of
purely hadronic decay channels on a qualitative level. Thus,
asymmetries, with the exception of the beam asymmetry, are
different in signs and absolute values.
Finally, we note that the spin-density matrices depend on the
choice of the reference frame (Gottfried-Jackson, or Helicity, or
Adair, cf.~\cite{TK07,Schilling70,TL03}). This circumstance must
be taken
into account in the analysis and interpretation of data.\\

  We thank H.~Schmieden for stimulating discussions.
  One of the authors (A.I.T.) appreciates the hospitality in FZD.
  This work was supported by BMBF grant 06DR136 and GSI-FE.


  \end{document}